\documentstyle[12pt,twoside]{article}
\input psfig.sty
\def\double12 {\smallskipamount=6pt plus2pt minus2pt
                  \medskipamount=12pt plus4pt minus4pt
                  \bigskipamount=24pt plus8pt minus8pt
                  \normalbaselineskip=24pt plus0pt minus0pt
                  \normallineskip=2pt
                  \normallineskiplimit=0pt
                  \jot=6pt
                  {\def\smallskip {\vskip\smallskipamount}}
                  {\def\medskip   {\vskip\medskipamount}}
                  {\def\bigskip   {\vskip\bigskipamount}}
                  {\setbox\strutbox=\hbox{\vrule
                    height17.0pt depth7.0pt width 0pt}}
                  \parskip 0pt
                  \normalbaselines}

\def\half12 {\smallskipamount=6pt plus2pt minus2pt
                  \medskipamount=12pt plus4pt minus4pt
                  \bigskipamount=24pt plus8pt minus8pt
                  \normalbaselineskip=16pt plus0pt minus0pt
                  \normallineskip=2pt
                  \normallineskiplimit=0pt
                  \jot=6pt
                  {\def\smallskip {\vskip\smallskipamount}}
                  {\def\medskip   {\vskip\medskipamount}}
                  {\def\bigskip   {\vskip\bigskipamount}}
                  {\setbox\strutbox=\hbox{\vrule
                    height17.0pt depth7.0pt width 0pt}}
                  \parskip 0pt
                  \normalbaselines}

\def\single12 {\smallskipamount=3pt plus2pt minus2pt
                  \medskipamount=6pt plus4pt minus4pt
                  \bigskipamount=12pt plus8pt minus8pt
                  \normalbaselineskip=12pt plus0pt minus0pt
                  \normallineskip=1pt
                  \normallineskiplimit=0pt
                  \jot=3pt
                  {\def\smallskip {\vskip\smallskipamount}}
                  {\def\medskip   {\vskip\medskipamount}}
                  {\def\bigskip   {\vskip\bigskipamount}}
                  {\setbox\strutbox=\hbox{\vrule
                    height8.5pt depth3.5pt width 0pt}}
                  \parskip 0pt
                  \normalbaselines}

\def\refitem{\par\noindent\hangindent 20pt}

\def\wisk#1{\ifmmode{#1}\else{$#1$}\fi}

\def\le     {\wisk{_<\atop^=}}

\def\deg    {\wisk{^\circ}}

\begin{document}
\pagestyle{plain}
\half12

\large
\begin{center}
Likelihood Analysis for Mega-Pixel Maps
\end{center}

\medskip
\normalsize
\half12
\noindent
\begin{center}
A.~Kogut\footnotemark[1]$^{,2}$
\end{center}
\footnotetext[1]{
~Laboratory for Astronomy and Solar Physics, 
Code 685, NASA/GSFC, Greenbelt MD 20771. \newline
\indent~$^2$ E-mail: Alan.Kogut@gsfc.nasa.gov. \newline
}

\medskip
\normalsize
\half12
\begin{center}
Accepted for publication by\\
{\it The Astrophysical Journal Letters} \\
\end{center}

\medskip
\begin{center}
\large
ABSTRACT
\end{center}

\normalsize
\noindent
The derivation of cosmological parameters from astrophysical data sets
routinely involves operations counts which scale as $O(N^3)$
where $N$ is the number of data points.
Currently planned missions, including MAP and Planck, will
generate sky maps with $N_d = 10^6$ or more pixels.
Simple ``brute force'' analysis,
applied to such mega-pixel data,
would require years of computing
even on the fastest computers.
We describe an algorithm which allows
estimation of the likelihood function
in the direct pixel basis.
The algorithm uses a conjugate gradient approach
to evaluate $\chi^2$
and a geometric approximation to evaluate the determinant.
Monte Carlo simulations
provide a correction to the determinant,
yielding an unbiased estimate of the likelihood surface
in an arbitrary region surrounding the likelihood peak.
The algorithm requires $O(N_d^{3/2})$ operations and $O(N_d)$ storage
for each likelihood evaluation,
and allows for significant parallel computation.

\noindent
{\it Subject headings:}
methods: data analysis --
cosmic microwave background

\clearpage
\section{Introduction}
Recent advances in instrumentation have transformed observational cosmology
from a field starved for data
to one where data management is fast becoming a limiting factor.
Prior to 1992, for example, there were {\it no} detections of anisotropy 
in the cosmic microwave background (CMB).
The Cosmic Background Explorer opened the floodgates
with maps containing $10^4$ pixels,
while scheduled missions such as MAP and Planck
plan for $10^6$ or more pixels.
While highly desirable from a scientific standpoint,
the explosive growth of cosmological data sets
carries the risk that their sheer size
will hamper analysis 
through computational limits on existing or planned computers.

Maximum likelihood methods are commonly used 
for parameter estimation 
with maps of the cosmic microwave background.
For a multivariate Gaussian distribution,
the probability of obtaining $N_d$ data points $\Delta_i$
given a set of model parameters $p$ is
\begin{equation}
{\cal L} = P(\Delta | p) =
(2\pi)^{-N_d/2} ~
\frac{ \exp( -\frac{1}{2} \chi^2 )}
     { |{\bf M}|^{1/2} }
\label{like_def}
\end{equation}
where 
\begin{equation}
\chi^2 = \sum_{i=1}^{N_d} \sum_{j=1}^{N_d}
\Delta_i ({\bf M}^{-1})_{ij} \Delta_j,
\label{chisq_def}
\end{equation}
is a goodness-of-fit statistic
and ${\bf M}_{ij} $
is the $N_d \times N_d$ covariance matrix.
The ``best'' choice of parameters $p_0$
is that which maximizes the likelihood function ${\cal L}$.
The curvature of the likelihood surface about the maximum
defines the uncertainty in the fitted parameters,
\begin{equation}
\delta p_j \geq \sqrt{ ({\bf F}^{-1})_{jj} }
\label{err_def}
\end{equation}
where
\begin{equation}
{\bf F}_{ij} = \langle
\frac{ \partial^2 L }
     { \partial p_i \partial p_j } \rangle
\label{fisher_def}
\end{equation}
is the Fisher information matrix
and $L = -\log({\cal L})$
(see 
Bunn \& Sugiyama 1995;
Vogeley \& Szalay 1996;
Tegmark et al.\ 1997;
Bond et al.\ 1998).

The maximum likelihood estimator is unbiased
and asymptotically approaches the equality in Eq. \ref{err_def}.
However, these advantages come at a steep price:
computation of both
$\chi^2$ and the determinant $|{\bf M}|$
in Eq. \ref{like_def}
scale as $O(N_d^3)$ operations,
making brute-force calculation computationally infeasible.
For large data sets ($N_d > 10^6$)
the time required is measured in years,
even on the most powerful computers.

Conjugate gradient techniques provide 
half of the solution.
We may reduce the operations count of the $\chi^2$ calculation
by computing the vector 
$$
z = {\bf M}^{-1} \Delta
$$
in Eq. \ref{chisq_def},
effectively trading the cost of a matrix inversion
for the requirement of re-calculating $z$
for every different sky map $\Delta$.
If the the matrix {\bf M} is dominated by the diagonal elements
(or is pre-conditioned to be suitably close to the identity matrix)
an iterative solution for $z$ can be found in $O(N_d^2)$ operations
(Press et al.\ 1992 and references therein;
Oh, Spergel, \& Hinshaw 1999).
The determinant calculation, however, is less tractable.
A number of authors have suggested ways around this problem.
Karhunen-Lo\`eve eigenvalue techniques
produce moderate data compression,
reducing the $N_d$ original data points
to $N^\prime \approx N_d/10$ eigenmodes
(Bond 1994;
Bunn \& Sugiyama 1995;
Tegmark et al.\ 1997).
However, 
estimating cosmological parameters from the smaller set of eigenmodes
still scales as $(N^\prime)^3$ operations, 
making such techniques undesirable for mega-pixel data sets.

Oh et al.\ (1999) derive a method for likelihood evaluation 
using a Newton-Raphson quadratic iteration scheme.
The determinant is first approximated 
using azimuthal symmetry of the noise matrix
(appropriate for full-sky CMB maps),
then corrected using Monte Carlo simulations.
The method provides a nearly minimum-variance estimate
of the angular power spectrum for CMB anisotropy maps
in $O(N_d^2)$ operations and $O(N_d^{3/2})$ storage;
cosmological parameters can then be derived
by comparing the power spectrum to various models.
Although this algorithm is fast enough for mega-pixel data sets,
it is optimized to estimate the power spectrum,
rather than the underlying cosmological parameters.
When used as a root-finding technique in parameter space,
it requires a sufficiently good starting estimate 
to guarantee convergence to true maximum.
The radius of convergence in parameter space is small
and the problems associated with parameter covariance become severe.

Borrill (1998) offers a global solution to bound the likelihood.
This method uses Gaussian quadrature to bound the likelihood
at {\it any} point in parameter space 
(not just near the likelihood maximum);
it is thus well suited 
to search parameter space using the direct pixel basis.
However, the method requires $O(N_d^{7/3})$ operations
for each likelihood evaluation
and is thus significantly slower than the method of  Oh et al.\ (1999).
More importantly, it can only provide bounds on the likelihood,
fixing $\log({\cal L})$ to accuracy of a few percent.
Since $\log({\cal L}) > N_d$ (a large number),
errors of a few percent can create significant bias
in the location of the likelihood maximum.

This paper describes an algorithm
to estimate cosmological parameters
from mega-pixel data sets
using maximum-likelihood techniques
on pixelized sky maps.
It uses a conjugate gradient algorithm
to evaluate $\chi^2$
and a geometric approximation to evaluate the determinant.
Monte Carlo simulations 
provide a correction to the determinant,
yielding an unbiased estimate of the likelihood surface
in an arbitrary region surrounding the likelihood peak.
The algorithm requires $O(N_d^{3/2})$ operations and $O(N_d)$ storage
for each likelihood evaluation,
and allows for significant parallel computation.

\section{Likelihood Evaluation}

Let the input data consist of a map of the microwave sky -- 
a vector consisting of temperature differences $\Delta_i$
evaluated at $N_d$ pixels on the sky.
The temperature in each pixel consists of a cosmological signal
plus instrument noise, 
$ \Delta_i = s_i + n_i$,
where we ignore for now the question of contaminating foreground signals
(galactic and extragalactic emission).
It is convenient to expand the CMB signal in spherical harmonics,
\begin{equation}
s(\Omega) = \sum_{\ell m} a_{\ell m} Y_{\ell m}(\Omega).
\label{harm_eq}
\end{equation}
Inflationary models predict the $a_{\ell m}$ to be 
random Gaussian variables 
whose variance depends only on angular scale $\ell$ and not on $m$.
The covariance matrix ${\bf M}_{ij}$ is thus
\begin{equation}
M_{ij}  = \langle \Delta_i \Delta_j \rangle
        = \frac{1}{4 \pi} \sum_\ell (2\ell + 1) W^2_\ell C_\ell
        P_\ell(\hat{n_i} \cdot \hat{n_j})
        + \frac{\sigma_0^2}{N_i^{\rm obs}} \delta_{ij},
\label{covar_eq}
\end{equation}
where $W^{2}_{\ell}$ is the experimental window function
that includes the effects of beam smoothing and finite pixel size,
$P_l(\hat{n_i} \cdot \hat{n_j})$
is the Legendre polynomial of order $\ell$,
$\hat{n_i}$ is the unit vector towards the center of pixel $i$,
$N_i^{\rm obs}$ is the number of observations for pixel $i$,
and $C_{\ell}$ is the angular power spectrum
$$
\langle a_{\ell m} a_{\ell^\prime m^\prime} \rangle
= C_{\ell} \delta_{\ell \ell^\prime} \delta_{m m^\prime}.
$$
We wish to evaluate the function
$$ 
L = -\log({\cal L}) \propto \chi^2 + \log(~|{\bf M}|~)
$$
to derive the set of parameters 
$p~=~[\Omega, \Omega_b, H_0, \Lambda, ...]$
which minimize $L$.

\begin{figure}[b]
\centerline{
\psfig{file=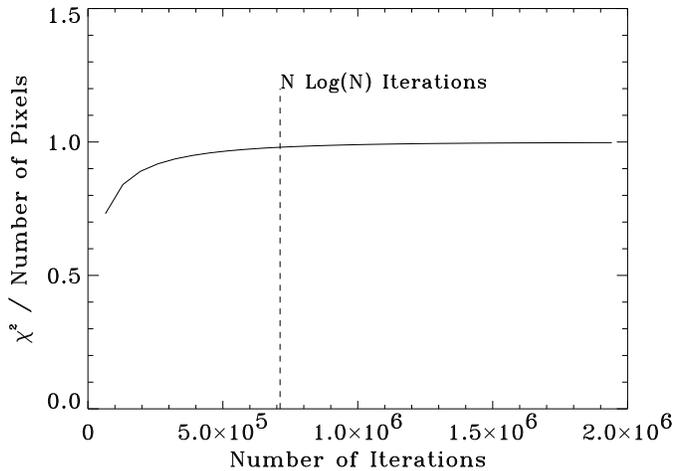,height=2.75in,angle=90}}
\caption{\single12
Convergence of $\chi^2$
evaluated using the conjugate-gradient algorithm
on a simulated CMB map with 64720 pixels.
\label{chisq_vs_iter}
}
\end{figure}

\subsection{Chi-Squared Evaluation}

Conjugate gradient techniques, to be efficient,
require the matrix ${\bf M}$ to be suitably close
to the identity matrix.
For CMB maps, this is almost always the case.
To see this, note that
${\bf M}_{ij} = \langle \Delta_i \Delta_j \rangle
= C(\theta_{ij})$
where
$C(\theta)$
is the two-point correlation function.
Since
$C(0) > C(\theta)$ for all $\theta > 0$,
the matrix must satisfy 
\begin{equation}
{\bf M}_{kk} > {\bf M}_{ik}, ~~~i \ne k
\label{max_k_eq}
\end{equation}
A particularly simple iterative method 
for the $\chi^2$ evaluation 
can be derived as follows.
We define a residual vector
\begin{equation}
r = {\bf M} z - \Delta
\label{resid_def}
\end{equation}
which by definition is zero when $z$ is solved exactly.
Now suppose that we have a guess which differs from the exact
solution by amount $\delta z$ in the $k^{\rm th}$ element, 
with corresponding residual vector 
$ r_i = \delta z ~ {\bf M}_{ik}$,
that is, the scalar $\delta z$
multiplied by the $k^{\rm th}$ row of the covariance matrix.
But from Eq. \ref{max_k_eq}
we may identify $k$ as the position of the largest element of the vector $r$,
and apply corrections
\begin{eqnarray}
z_k = z_k - \delta z \nonumber \\
r_i = r_i - \delta z ~{\bf M}_{ik}
\label{correction_eq}
\end{eqnarray}
where
\begin{equation}
\delta z = {\rm Max}(R) ~/ ~{\bf M}_{kk}
\label{delta_z_def}
\end{equation}
For the general case where more than one element of $z$ is incorrect,
we may proceed iteratively,
using Eqs. \ref{delta_z_def} and \ref{correction_eq}
until the largest element in the residual vector
falls below some pre-determined threshold.

Figure \ref{chisq_vs_iter}
shows the convergence of $\chi^2$ 
for a simulated sky map
and vector $z$
evaluated using the algorithm above
for the case where ${\bf M}$
correctly describes the input map.
The convergence is rapid.
Each iteration requires $O(N_d)$ operations
(the vector multiplication in Eq. \ref{correction_eq})
while the entire algorithm converges in $N_d \log(N_d)$ iterations.

\subsection{Determinant Approximation}

The determinant calculation in Eq. \ref{like_def}
scales as $O(N_d^3)$ operations,
making direct evaluation prohibitively expensive.
We make a first approximation to $| {\bf M} |$ 
by noting that
the elements ${\bf M}_{ij}$ depend only on the angular separation
between pixels $i$ and $j$.
For a rotationally invariant signal,
rows ${\bf M}_{ik}$ and ${\bf M}_{jk}$
thus consist of the (nearly) the same numbers, 
simply repeated in different order
according to the pixelization.
All of the information in the covariance matrix
is contained in the two-point correlation function $C(\theta)$ --
the covariance matrix merely samples $C(\theta)$
at a set of discrete values,
then orders the values according to the pixel scheme.
We thus make the {\it ansatz} that
\begin{equation}
\log( | {\bf M} | ) = \alpha \log( | {\bf M^{\prime}} | )
\label{ansatz_eq}
\end{equation}
where the covariance matrix ${\bf M^{\prime}}$
is formed from a subset of the pixels in the original sky map,
chosen to sample the correlation function
with the same angular distribution
as the full set of pixels.
If there are  $N^{\prime}$ pixels in the subset,
computing $| {\bf M^{\prime}} |$ 
costs $(N^\prime)^3$ operations.
We thus restrict ourselves to
$N^\prime \le N_d^{2/3}$
so that the entire calculation scales as $N_d^2$
or faster.

The subset of $N^{\prime}$ pixels 
from the original sky map
can be selected several ways.
A random set of pixels will,
on average,
match the distribution of angular separations $\theta_{ij}$
in the original sky map.
However, sample variance becomes important for small subsets
and can over- or under-represent some separations
in a single realization.
While this would average out over many realizations,
we are only allowed a single realization
in the likelihood evaluation.
We can overcome this by
using subset pixels in a fixed geometric orientation.
For full-sky maps, 
we obtain satisfactory results
using a set of three interlocking great circles.
Other configurations can be devised
for different map geometries.
Once the subset of $N^\prime$ pixels is selected, 
the determinant $|{\bf M^\prime}|$
can be computed using standard techniques.
For the great-circle geometry,
the number of subset pixels scales
as $N^\prime \propto \sqrt{N_d}$.
The determinant $|{\bf M^\prime}|$
thus requires $O(N_d^{3/2})$ operations
and $O(N_d)$ storage.

To derive the scale factor $\alpha$,
we again take advantage of the fact that 
for CMB maps the matrices ${\bf M}$ and ${\bf M^\prime}$ 
are suitably close to diagonal.
For diagonal matrices,
\begin{equation}
\alpha_0 = \frac{ \sum_i \log( {\bf M}_{ii} ) }
	      { \sum_i \log( {\bf M^\prime}_{ii} ) }.
\label{alpha_approx}
\end{equation}
Off-diagonal elements produce corrections to this simple relation.
For the CMB maps expected from the MAP and Planck surveys,
however,
Equation \ref{alpha_approx} is accurate to a few percent.
We have tested this simple scaling for
maps ranging from 384 to 6144 pixels,
the largest value for which the full determinant can readily be evaluated.
The simplistic approximation
is accurate to a few percent over a wide range of parameter space
and map size.

\subsection{Monte Carlo Correction}

Equations \ref{ansatz_eq} and \ref{alpha_approx}
allow efficient evaluation of the logarithmic likelihood 
\begin{equation}
L = \chi^2 + \alpha |{\bf M^\prime}|
\label{like_approx}
\end{equation}
at any point in parameter space to an accuracy of $\sim$2\%
for computational cost $O(N_d^{3/2})$ and storage $O(N_d)$.
It is thus well-suited for the initial stages
of parameter estimation in the direct pixel basis.
In this scheme,
we begin with some initial choice of parameters
and use the methods outlined above
to evaluate the likelihood in the neighborhood
of the initial parameters.
Using established minimization search techniques
(see, e.g., Press et al.\ 1992)
we then step through parameter space toward the global likelihood maximum.

The simple approximation to $L$
differs from the true value by a few percent 
in a smoothly-varying fashion.
It thus returns a biased estimate of the global maximum,
but does not create secondary (local) maxima.
However, the initial stages of a parameter search
overwhelmingly involve the rejection of {\it bad} solutions
for the parameter values in favor of better ones;
as such, we can afford to sacrifice 
some initial accuracy for computational speed.
Once we have found the approximate location of the global maximum,
we use Monte Carlo simulations
to iterate toward an exact solution for the likelihood surface
in a much smaller region of parameter space.

\begin{figure}[b]
\centerline{
\psfig{file=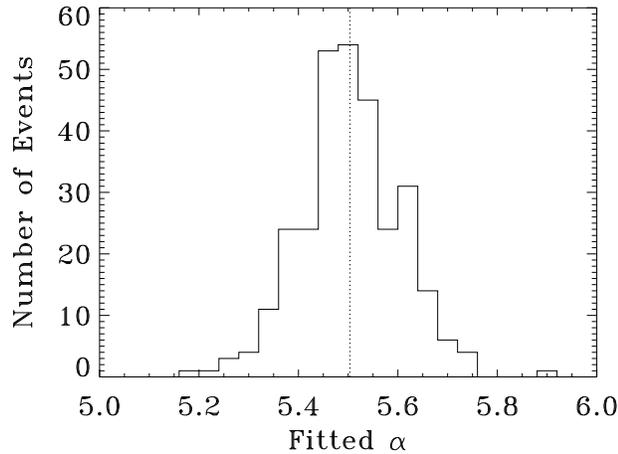,height=2.75in,angle=90}}
\caption{\single12 Histogram of the scale factor $\alpha$
derived from Eq. \ref{like_approx}
using 300 Monte Carlo simulations.
The dotted line indicates the value from direct evaluation
of the determinants $|{\bf M}|$ and $|{\bf M^\prime}|$.
The simulations shown correspond to input parameters
$\Omega=0.7$, $n=1.0$ and normalization $Q=18 ~\mu$K,
but the ability to recover the correct scaling $\alpha$
does not depend on the location in parameter space.}
\label{alpha_histogram}
\end{figure}

Monte Carlo techniques reverse the sense of the likelihood evaluation:
instead of using Eqs. \ref{alpha_approx}
and \ref{like_approx} 
to evaluate the likelihood for a map 
with unknown cosmological parameters,
we use a set of simulated maps 
generated from a single fixed choice of parameter values
and solve Eq. \ref{like_approx} for the unknown value $\alpha$
for which the likelihood 
correctly peaks at the known parameter values.
That is, we generate a set of $m$ simulated sky maps
from a single set of parameter values $p_0$.
For each map, we evaluate $\chi^2$ 
at a grid of parameter values 
centered on the input parameters 
(in practice, we find that varying a single parameter
by a few percent provides acceptable results).
We then use the $\chi^2$ grid
and determinants $|{\bf M^\prime}|$
(which do not depend on the map values)
to solve Eq. \ref{like_approx} 
for the scale factor $\alpha$
by requiring that $\log({\cal L})$
peak at the input parameter value.

The scale factor $\alpha$ is a weak function of the parameter values.
The fitted value $\alpha$ from each simulation
is thus exact only at the parameter values $p_0$,
becoming increasingly poor
at grid points further removed from $p_0$.
However,
since the simulated maps correspond to the parameter choice $p_0$,
the likelihood function must peak exactly at $p_0$
and the process by construction
is quadratically convergent.

We test the algorithm using simulated sky maps
with 6144 pixels
using different choices for
sky cut, instrument noise, and subset pixel selection.
We generate each simulated map
using Eq. \ref{harm_eq}
with random variables $a_{\ell m}$
corresponding to a COBE-normalized cold dark matter model
with $\ell < 100$ and $3\deg$ beam smoothing.
To each sky we add instrument noise
following either the DMR noise pattern
(Bennett et al.\ 1996)
or spatially uniform noise
of the same mean amplitude.
For each realization
we compute the determinant $|{\bf M^\prime}|$
and evaluate $\alpha$ 
using both the great circle and random pixel geometries
on both the full sky map
or the high-latitude portion $|b| > 20\deg$.

Figure \ref{alpha_histogram}
shows a histogram of the recovered values
from $m=300$ Monte Carlo realizations
for the full sky map with uniform noise and the great circle geometry.
The recovered value,
$\alpha = 5.504 \pm 0.006$,
agrees well with the exact value (5.5036)
derived by direct evaluation of
the determinants $|{\bf M}|$ and $|{\bf M^\prime}|$.
Since the method involves no integrals over spatial functions
it is completely insensitive to sky coverage.
Allowing the noise to vary with position on the sky
increased the scatter in the recovered values for $\alpha$ 
but did not bias the results:
300 simulations using COBE noise
with $|b| > 20\deg$ and random pixels
obtained $\alpha = 7.07 \pm 0.04$
compared to the exact value 7.02
for 4016 high-latitude pixels.
The convergence is rapid:
One round of Monte Carlo realizations
usually suffices to derive an accurate likelihood
for a single point in parameter space.

\section{Discussion}

The determinant approximation
requires $O(N_d^{3/2})$ operations and $O(N_d)$ storage
for each point evaluated
in parameter space.
Conjugate gradient evaluation of $\chi^2$ requires
$O( m ~N_d^2 )$ operations,
where $m$ is the number of Monte Carlo realizations.
During the initial parameter search,
Monte Carlo simulations are turned off and $m=1$.
Once an initial maximum is located,
$m \approx 50$ realizations
locate the true maximum to within 1\% 
of the minimum variance limit.
With efficient pre-conditioning 
(Oh et al.\ 1999),
100 $\chi^2$ evaluations for a $10^6$ pixel map
can be run in 2 hours elapsed time.
Since the determinant depends only on the model parameters
and not on the map temperatures in each realization,
it need be computed only once for each set of parameter values.
Timing tests using a single CPU at 400 MHZ
demonstrate that the total elapsed time 
required to calculate $|{\bf M^\prime}|$
scales as $N_d^{1.55}$
including all overhead.
Computing the determinant for a map with $N_d = 10^6$ pixels
requires 24 hours elapsed time on a single CPU.
The entire computation thus has
one term that scales as
$\Delta t ~\sim ~2 ~{\rm hours} ~(N_d / 10^6)^2$
and a second term scaling as
$\Delta t ~\sim ~24 ~{\rm hours} ~(N_d / 10^6)^{3/2}$.
For $N_d < 10^8$ pixels,
the determinant is the slower of the two
and the entire computation scales as $N_d^{3/2}$.

The likelihood method described above has several advantages.
By construction,
the Monte Carlo simulations converge to the correct solution.
Since the Monte Carlo algorithm evaluates the likelihood
at parameter values close to the peak likelihood,
it automatically provides the local curvature 
(Fisher information matrix)
at no additional computational cost.
The technique is well suited for the modest parallel processing
provided by multiple-CPU work stations currently available.
Finally, the technique does {\it not} depend on assumptions
of any symmetry on the sky
and can be used for any sky map
regardless of the shape of the observed region.
Asymmetric maps
(e.g., a cut map that excludes bright regions/point sources
or a partial map from balloon surveys)
may be evaluated using the techniques outlined above.

\vspace{0.5 in}
Gary Hinshaw and David Spergel provided helpful comments.
This work was funded in part by NASA grant S-57778-F.

\bigskip
\bigskip
\begin{center}
\large
{\bf References}
\end{center}

\refitem
Bennett, C.L., et al.\ 1996, ApJL, 464, L1

\refitem
Bond, J.R.\ 1994, 
PRL, 72, 13

\refitem
Bond, J.R., Jaffe, A.H., and Knox, L.\ 1998, PRD, 57, 2117

\refitem
Borrill, J.\ 1998, Phys. Rev. (submitted); 
preprint astro-ph/9803114

\refitem
Bunn, E.F., and Sugiyama, N.\ 1995,
ApJ, 446, 49

\refitem
Oh, S.P., Spergel, D.N., and Hinshaw, G.\ 1999, ApJ, 510, 551

\refitem
Press, W.H., Teukolsky, S.A., Vetterling, W.T., and Flannery, B.P.\ 1992,
Numerical Recipes (Cambridge University Press: Cambridge)

\refitem
Tegmark, M., Taylor, A.N., and Heavens, A.F.\ 1997,
ApJ, 480, 22

\refitem
Vogeley, M.S., and Szalay, A.S.\ 1996,
ApJ, 465, 34

\end{document}